\documentclass{pasj01}
\usepackage[switch]{lineno}
\Received{$\langle$reception date$\rangle$}
\Accepted{$\langle$acception date$\rangle$}
\Published{$\langle$publication date$\rangle$}

\begin{document}

\title{SN 2024iss: Double-Peaked Light Curves and Implications for a Yellow Supergiant Progenitor}
\author{Masayuki Yamanaka$^{1}$, Takahiro Nagayama$^{2}$, and Tsukiha Horikiri$^{3}$}
\altaffiltext{1}{Amanogawa Galaxy Astronomy Research Center (AGARC), Graduate School of Science and Engineering, Kagoshima University, 1-21-35 Korimoto, Kagoshima, Kagoshima 890-0065, Japan}
\altaffiltext{2}{Graduate School of Science and Engineering, Kagoshima University, 1-21-35 Korimoto, Kagoshima, Kagoshima
890-0065, Japan}
\altaffiltext{3}{Department of Physics and Astronomy, Faculty of Science, Kagoshima University, 1-21-35 Korimoto,
Kagoshima, Kagoshima 890-0065, Japan}
\email{yamanaka@sci.kagoshima-u.ac.jp}


\maketitle

\begin{abstract}
 We report the multi-band photometric observations of the Type IIb supernova (SN) 
 2024iss with ultra-violet (UV), optical, and near-infrared (NIR) wavelengths 
 starting one day after the explosion. The UV and optical light curves show the first peak two days after the explosion date. Following a first peak, 
 a secondary maximum is observed in the optical and NIR bands, similar to SNe IIb with double-peaked light curves. 
 The quasi-bolometric light curve shows the fast decay until a week after the explosion. 
 From the analysis of the bolometric light curve, the ejecta mass and kinetic energy are estimated to be $M_{ej}=2.8\pm0.6~M_{\odot}$ and $E_{kin}=9.4\pm4.1\times10^{50}$ erg. The 
 mass of the radioactive $^{56}$Ni is estimated to be $M(^{56}Ni)=0.2~M_{\odot}$.
 Fitting a black-body function to the spectral energy distribution reveals that the photospheric temperature exhibits a rapid exponential decline during the first week after the explosion.
 An analytic model describing the cooling emission after shock breakout provides a reasonable explanation for the observed temperature evolution.
 From these ejecta parameters, we calculated the progenitor radius to be $R_{pro}=50-340$~$R_{\odot}$. We conclude that these explosion properties are consistent with a core-collapse explosion from a yellow supergiant (YSG) progenitor.  
\end{abstract}

\section{Introduction}
  Core-collapse supernovae (CC SNe) occur at the final stage of the stellar evolution of the massive star. Observational properties of 
  some CC SNe are explained by the theoretical explosion scenario for the stripped-envelope progenitor \citep{Nomoto1993,Woosley1994}. 
 The progenitors of Type IIb SNe have a thin hydrogen envelope,
 although it is quite ambiguous how the progenitor strips off the envelope through the stellar evolution \citep{Smith2011,Sana2012}.
 Observational research of Type IIb SNe could provide an important key to approaching this problem.
 
 The cooling emission has been observed in the light curves of Type IIb SNe 1993J \citep{Richmond1994}, 2011fu \citep{Kumar2013, Morales-Garoffolo2015}, 2013df \citep{VanDyk2014,Morales-Garoffolo2014,Szalai2016}, 2016gkg 
 \citep{Arcavi2017,Tartaglia2017}, and 2020bio \citep{Pellegrino2023} in optical and ultra-violet (UV) wavelengths.
 The emission comes from the cooling envelope after the shock breakout. After the cooling phase, the secondary maximum is normally observed.
 The photospheric temperature is higher in earlier phases, and it rapidly decreases. 
 From the analysis of the cooling emission \citep{Morales-Garoffolo2014,Tartaglia2017,Pellegrino2023} and the pre-explosion images \citep{Aldering1994,Crockett2008,VanDyk2014}, the progenitor is proposed as a yellow supergiant (YSG) star.
 However, there are no comprehensive 
 photometric analyses including the near-infrared (NIR) data up to date. 
 
  SN 2024iss was discovered at 14.6 magnitude by the 
  Gravitational-wave Optical Transient Observer (GOTO; \cite{Steeghs2022}) 
  on May 12.9 in 2024 (UT) \citep{ONeill2024}. This SN was identified
  as a typical Type IIb SN by two-epoch 
  spectroscopic observations \citep{Srivastav2024b}.
  The last non-detection magnitude
  was obtained as 19.5 mag on May 11.9 by the GOTO 
  collaboration \citep{ONeill2024}. It means that the light curve shows
  a rapid rise during a day. From this constraint, we define that
  the breakout time ($t=0$ d) of this object was at 
  May 12.4 (UT) with an uncertainty of $\sim0.5$ day.

   The host galaxy of SN 2024iss is faint and compact \citep{Godwin1983}. 
  The uncertainty of the distance is very large.
  We adopted the distance of 14.1 Mpc, converted
  by the redshift $z=0.003334$ reported by \citep{ONeill2024}.
  The corresponding distance modulus, $\mu=30.7$. In 
  the case of the large distance (19.7~Mpc), the peak absolute 
  magnitude of SN 2024iss was -18.2 mag in $g$-band, which indicates 
  the very luminous for SN IIb. In this Letter, we 
  adopted the former distance for the conservative estimation 
  of the luminosity. We also adopted a Galactic extinction of $A_{V}=0.027$ \citep{Schlafly2011} throughout this Letter.

  In this Letter we will present the observations and 
  data reduction in \S 2, the photometric results
  and comparison of the quasi-bolometric light curves 
  with other objects in \S 3, and the analysis of the 
  spectral energy distribution and the evolution of the 
  photospheric temperature in \S 4. Finally, we present 
  the conclusion in \S 5.
   
\section{Observations \& data reduction}

  We simultaneously obtained the NIR and optical imaging data
 on eight nights from May 15.5 to Jun 13.6 in the $g$, $i$, $J$, 
 $H$, and $K_{s}$ bands with the kSIRIUS+gi camera 
 \citep{Nagayama2024} attached to the Kagoshima 1-m telescope 
 at the Iriki Observatory. 

  We carried out the data reduction through
 the standard manner using $IRAF$ \citep{Tody1986,Tody1993}. After that,
 we performed the point spread function (PSF)
 photometry of SN 2024iss and the reference stars using the 
 $DAOPHOT$ package \citep{Stetson1987}.
 We performed the photometric calibration
 of these data using the Pan-STARRS (PS1; \cite{Chambers2016}) and 
 Two Micron All Sky Survey (2MASS; \cite{Cutri2003}) catalogs. 

 We also used the public $g$ and $r$-band photometric
 data from the Zwicky Transient Facility (ZTF; \cite{Bellm2019}) survey, and we took the data from 
 the Automatic Learning for the Rapid Classification of 
 Events (ALeRCE; \cite{ZTF2021}) site.
 The first data point was obtained at nine hours 
 after the discovery on May 13.4 \citep{PFournon2024}. 
 We confirmed the consistency between the Kagoshima and 
 ZTF $g$-band light curves, indicating negligible 
 contamination from the host galaxy. We did not apply 
 template subtraction to the photometric data.
  We also performed the aperture photometry of the public
 optical and UV imaging data, obtained
 on 19 epochs from May 13.6 to Jun 23.0 using the UVOT at
 the Neil Gehrels Swift Observatory \citep{Roming2005}
 in $uvw2$, $uvm2$, $uvw1$, $U$, $B$, and $V$ -band. 
 We adopted the zeropoints and calibration database from 
 \citet{Breeveld2011}. Template subtraction was not performed 
 for the UVOT data.
 The multi-band light curves are shown in Figure 1.

\section{Results}

 In the UV and optical wavelengths, 
 the $uvw2$, $uvm2$, $uvw1$, $U$, $g$, $B$, $V$, and $r$-band light curves exhibit the first peak around $t\sim2$ d. After 
 that, the secondary maximum is found 
 at $t\sim17$ d in the $g$ and $r$ bands, and 
 at $t\sim22$ d 
 in the $K_{s}$-band. The longer-wavelength 
 band light curves reach the maximum date later.
 This fact is consistent with the trend seen in the normal 
 stripped envelope SNe \citep{Bianco2014,Ergon2014}.
  
\begin{figure}
    \includegraphics[width=0.48\textwidth]{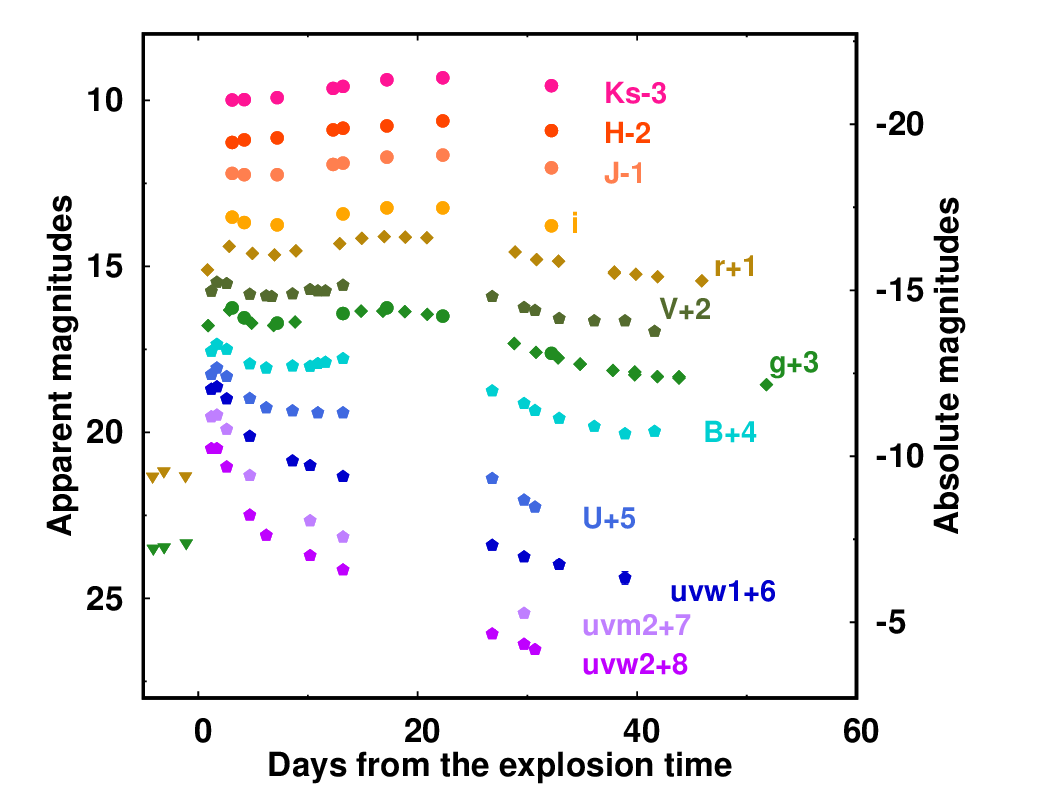}
     \caption{The UV, optical, and near-infrared 
    light curves of SN 2024iss. The left-side vertical 
    axis denotes the apparent magnitude and the right-side
    one does the absolute magnitude. The distance modulus
    was adopted for $\mu=30.7$ which was converted from the redshift of z=0.0033 \citep{ONeill2024}. The 
    filled-circle symbols denote the photometric point obtained using the kSIRIUS+gi camera. 
    The pentagon-shape symbol denotes data obtained by 
    UVOT. The diamond-shape symbol denotes data obtained by ZTF. The downward triangle denotes the upper-limit 
    magnitude obtained by ZTF. The $uvw2$, $uvm2$, $uvw1$, 
    $U$, $B$, $V$, $J$, $H$, and $K_{s}$-band magnitudes
    are in the Vega system, and $g$, $r$, and $i$-band
    ones in the AB system.
    {Alt text: The horizontal axis is scaled in days from the explosion.} }
\end{figure}

 We present the procedure of constructing the quasi-bolometric light curve and analyzing the emission at its first peak. 
 We converted the observed magnitudes into flux densities using the conversion factors provided by \citet{Fukugita1996}, \citet{Bessell1998} and \citet{Tokunaga2005}, and obtained 
 the spectral energy distribution.
 Using the central wavelength of each 
 passband function \citep{Fukugita1996,Bessell1990}, 
 we calculated the quasi-bolometric light curves using the distance of 14.1 Mpc.
 The Galactic extinction was corrected for \citep{Schlafly2011}. 
 The extinction in the host galaxy is negligible, as indicated by the non-detection of sodium absorption lines in the spectra \citep{Srivastav2024a,Srivastav2024b}. 
 We plotted the quasi-bolometric light curve of SN 2024iss in Figure 2.
 The light curves were compared with those of other Type IIb SNe \citep{Richmond1994,Marion2014,Morales-Garoffolo2014,Morales-Garoffolo2015,Tartaglia2017}.

 Light curve properties were extracted by fitting the light curves with a combined second-order polynomial and exponential function. The fit accurately describes the observed data. The uncertainty is primarily attributed to the distance. The peak luminosity of SN 2024iss is determined to be $5.1\times10^{42}$ erg~s$^{-1}$ at $t=3.5$ d. 
  Thereafter, the luminosity exhibited a rapid decline to $2.5\times10^{42}$ erg~s$^{-1}$ at $t=7.6$ d. We defined 
 as the cooling phase until this epoch. The light curve evolution was 
 nearly flat. The first peak luminosity is higher than 
 that of SNe 1993J and 2016gkg during  the comparable evolutionary 
 phases. 
 We will present the analysis of the progenitor properties using the light curve in \S 4.2. 
 
 The secondary maximum was marginally detected, with its luminosity of $\sim2.6\times10^{42}$ erg~s$^{-1}$ at $t=15.3$ d.
  The secondary maximum was also much higher than those 
 of other objects, and the rise time is quite shorter than 
 those of the objects. From these properties, we will
 discuss the ejecta properties of this object in \S 4.1.

  We calculated the temporal evolution of the fraction of the 
UV and the NIR luminosity fraction against the bolometric light curve. 
The UV emission fraction shows a rapid decline until $t=7$ d from $~\sim40~\%$ to $~\sim30~\%$.
Thereafter, it linearly declines to below $\sim10~\%$ at $t=23$ d. 
The NIR emission fraction shows a rapid rise to the $20~\%$ 
until $t=7$ d, and it is 
larger than that of UV after $t=10$ d. Finally, the NIR emission 
reaches $\sim40~\%$ around $t=30$ d. This means that the NIR 
component is essentially important for the bolometric correction
after the cooling phase before the radioactive peak.

\begin{figure}
    \includegraphics[width=0.45\textwidth]{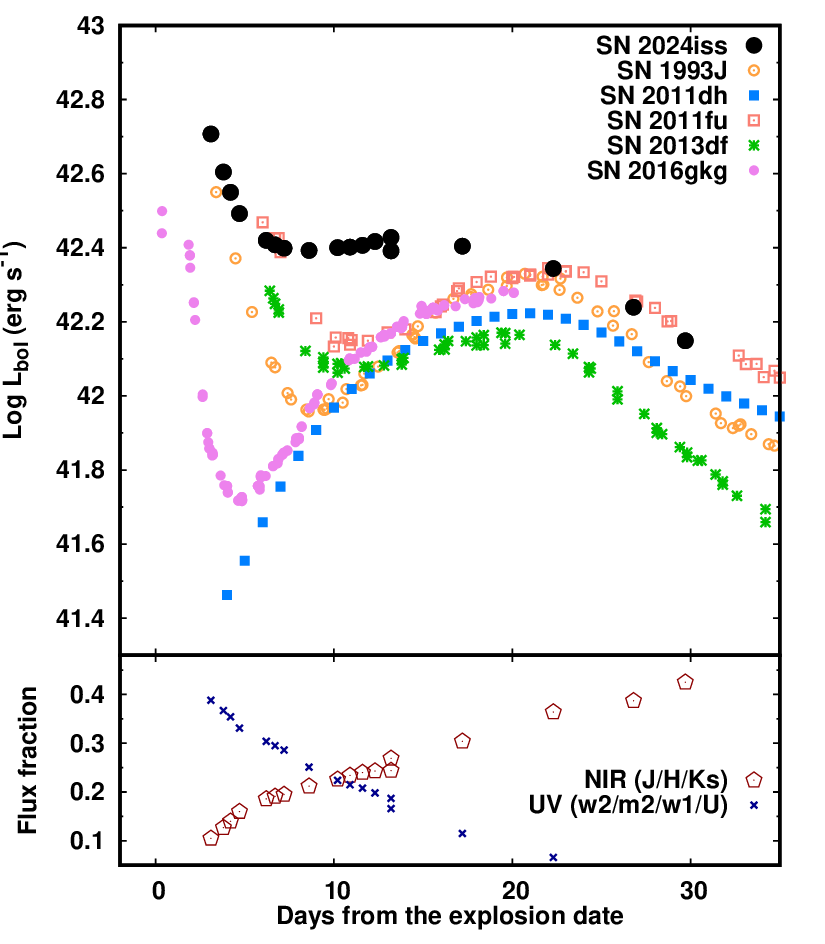} 
    \caption{(Upper panel) The quasi-bolometric light curve of SN 2024iss compared
    with those of SNe 1993J \citep{Richmond1994}, 2011dh \citep{Marion2014}, 2011fu \citep{Morales-Garoffolo2015}, 2013df \citep{Morales-Garoffolo2014} and 2016gkg \citep{Tartaglia2017}. The integration
    of each band flux was performed using the central wavelength of 
    each band. The Galactic extinction and the distance were corrected for. (Lower panel) The evolution of the UV and NIR fraction. The UV fraction is denoted by cross-shaped symbols in blue, and NIR is denoted by open pentagons in red.
    {Alt text: In the upper panel, the vertical axis shows the quasi-bolometric luminosity in logarithmic scale, in units of erg per second. In the lower panel, the vertical axis shows the fraction of the UV and NIR band flux relative to the total luminosity. 
    The horizontal axis is scaled in days from the explosion.}
    }
\end{figure}

 We classify SN 2024iss based on the publicly available spectrum obtained by \citet{Srivastav2024b}. The spectra are shown in Figure 3.
 The blueshifted absorption lines of H$\alpha$ and He~{\sc i} $\lambda$5876 are identified in the spectrum.
Comparing this spectrum with those of the well-studied Type IIb SNe 1993J \citep{Filippenko1993} and 2011dh \citep{Marion2014}, we find similar absorption features 
of hydrogen and helium. This strongly suggests that SN 2024iss is also a Type IIb. The measured expansion velocity of the H$\alpha$ line is 15,000 km~s$^{-1}$, which is consistent with the typical velocities observed in Type IIb SNe 
\citep{Pastorello2008,Marion2014,Medler2022}.

\begin{figure}
    \includegraphics[width=0.45\textwidth]{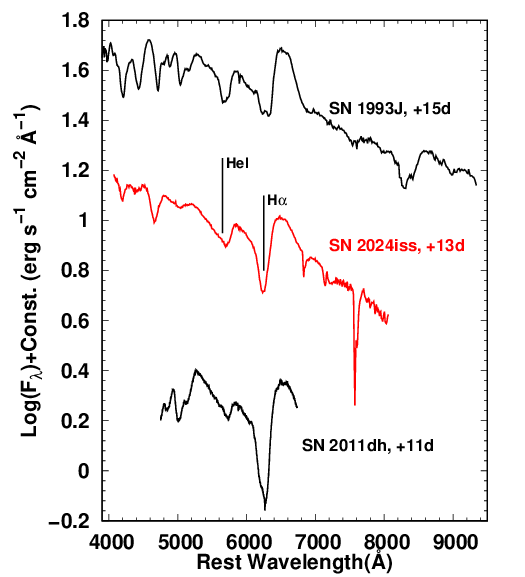} 
    \caption{Optical spectrum of SN 2024iss observed by \citet{Srivastav2024b}. 
    It is compared with that of
    SNe 1993J \citep{Filippenko1993} and SNe 2011dh \citep{Marion2014}.
    The wavelength of the host galaxy was corrected to
    the rest frame using the redshift presented by each literature.
    The absorption lines of H${\alpha}$ and He~{\sc i} are identified, 
    indicating that SN 2024iss is classified as a Type IIb.
    {Alt text: The vertical axis shows the flux in logarithmic scale, in units of erg per second per square centimeter per Angstrom. The horizontal axis shows the wavelength in units of Angstrom.}}
\end{figure}

\section{Discussion}


\subsection{Ejecta properties}
 The ejecta mass and kinetic energy can be estimated based on the 
 light curve timescale and the expansion velocity, using
 the relation described by \citet{Arnett1982}. 
 The rise time, representing the timescale to the peak, was calculated to be $t_{rise} = 15.3$ d, 
 by fitting the function described in \S 3 to the rising part in the bolometric light curve.
 We used the H${\alpha}$ line velocity of 15000 km~s$^{-1}$ from the 
 measurement presented in \S 3. For the conservative estimate, we adopted a 20\% error for these measurements.
  
  The scaling-law method \citep{Arnett1982} was 
  employed to obtain the 
 kinetic energy ($E_{kin}$) and the ejecta mass ($M_{ej}$)
 of SN 2024iss. We used the kinetic energy $E_{kin}=8.2\times10^{50}$ erg and $M_{ej}=2.7~M_{\odot}$ of SN 2011dh \citep{Bersten2012} as reference values to obtain 
 properties of SN 2024iss. 
 The line velocity of SN 2011dh was $v$(H${\alpha}$)$=13000$ km~s$^{-1}$ around at maximum, and the rise time of SN 2011dh was $t_{rise}=22$ d \citep{Marion2014}. 
 From these parameters, we derived the kinetic energy of 
 $E_{kin}=9.4\pm4.1\times10^{50}$ erg and ejecta mass of 
 $M_{ej}=2.8\pm0.6~M_{\odot}$. The uncertainty in the rise time 
 and line velocity of SN 2024iss contributes to the error.
 We will use these parameters
 for an analysis of the progenitor in \S 4.2.

  The mass of the radioactive $^{56}$Ni is dependent on a peak luminosity 
 and a rise time of the bolometric light curve \citep{Arnett1982}.
 In the case of Type IIb SNe with a double-peaked light curve,
 the secondary maximum can be explained by the presence of $^{56}$Ni.
 Based on the secondary peak luminosity of $\sim2.6\times10^{42}$ erg s$^{-1}$ and 
 the rise time of $t=15.3$ d, 
 the radioactive nickel mass of SN2024iss is estimated to be $M(^{56}Ni)=0.2$~$M_{\odot}$. 
 This value is consistent with the range of $^{56}$Ni masses observed in Type IIb SNe \citep{Lyman2016}.
 
\subsection{Analysis of the cooling phase}
 We constructed the spectral energy distribution (SED) of 
 SN2024iss for the seven epochs in the UV, optical, and NIR wavelengths (see Figure 4).
 The central wavelengths of the
 $uvw2$, $uvm2$, $uvw1$, $U$, $B$, $V$, $g$, $i$, $J$, $H$, and 
 $K_{s}$-band data were used in order to convert 
 the magnitudes to the flux density. We performed the 
 black-body fitting to the SED using the least squares method. 
 The estimated photospheric 
 temperature and radius are shown in Figure 5. While the temperature 
 rapidly decreases with time, the photospheric radius gradually increases.

 We also attempted to fit the theoretical model to the evolution of the temperature. 
 We used the blue supergiant (BSG) progenitor model from \citet{NS2010} and the yellow supergiant (YSG) progenitor model from \citet{Milisavljevic2013}, the latter of which was originally proposed by \citet{NS2010}. In these models, the temperature evolution depends on the ejecta mass, the ejecta kinetic energy, and the progenitor radius. The temperature evolution was well explained by both models within the uncertainty. From the BSG model, we estimated the progenitor radius to be, $R=340\pm40 R_{\odot}$ while from the YSG model, we estimated it to be $R=50\pm5 R_{\odot}$. Based on these fits, we conclude that the progenitor radius is likely $50-340 R_{\odot}$.

   For other SNe IIb, the progenitor radii were estimated to be 
 $\sim450~R_{\odot}$ for SN 2011fu 
 \citep{Morales-Garoffolo2015}, $64-169~R_{\odot}$ for SN 2013df 
 \citep{Morales-Garoffolo2014}, and $48-124~R_{\odot}$ for SN 2016gkg \citep{Tartaglia2017}
 using the similar method. 
 Our estimated progenitor size is consistent with that of SNe 2013df
 and 2016gkg, but smaller than that of SN 2011fu. 
 Considering the systematic difference of the model, the progenitor size
 of SN 2024iss was consistent with the range of Type IIb SNe. 
   
\begin{figure}
    \includegraphics[width=0.45\textwidth]{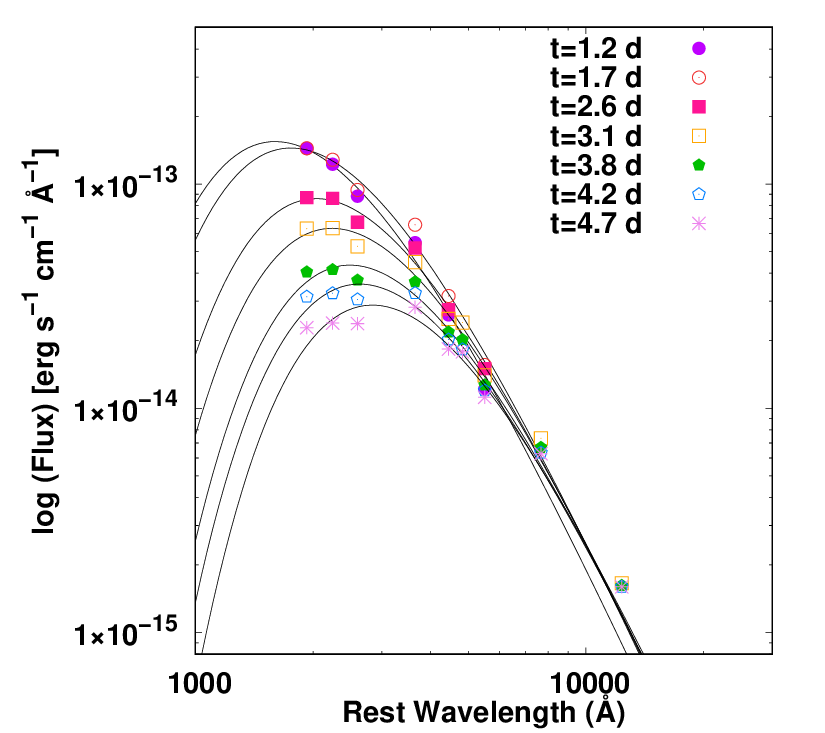}
    \caption{The spectral energy distribution of SN 2024iss
    during a week after the explosion between $t=1.2$ and $4.7$ d. 
    the black curves denote the black-body function.
    {Alt text: 
    The vertical axis shows the flux in logarithmic scale, in units of erg per second per square centi meter per Angstrom. The horizontal axis shows the wavelength in units of Angstrom. 
    The evolution of the spectral energy distribution is shown, with 
    each phase is represented by different colors. }}
\end{figure}

\begin{figure}
    \includegraphics[width=0.45\textwidth]{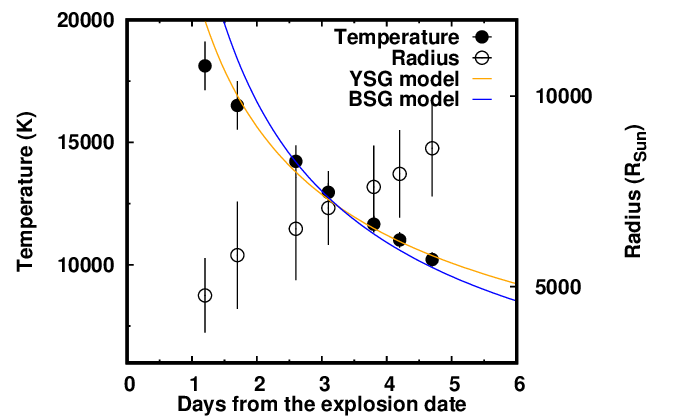}
     \caption{Evolution of photospheric temperature and radius of SN 2024iss.
     Black-color filled circle denotes the 
     evolution of the photospheric temperature. 
     The orange curve shows the best-fit solution for the YSG model proposed by 
     \citet{Milisavljevic2013}, while the blue curve shows that for 
     the BSG model \citep{NS2010}.
     The black-color open circle denote the temporal evolution
     of the photospheric radius.
     {Alt text: The left-side vertical axis shows the temperature in units of kelvin. The right-side axis shows the radius in units of 
     solar radius.}}
\end{figure}

\section{Conclusion} 
 We present the multi-band light curves of SN 2024iss 
 obtained in UV, optical, and NIR wavelengths simultaneously. 
 From the quasi-bolometric light curve, we estimated the 
 light-curve properties as follows. 
 The radioactive $^{56}$Ni mass was $0.2~M_{\odot}$,
 the total ejecta mass was $M_{ej}=2.8\pm0.6~M_{\odot}$, and
 the kinetic energy of ejecta was $E_{kin}=9.4\pm4.1\times10^{50}$ erg.
 These ejecta parameters are consistent with those of comparative Type IIb SNe.
 From the comparison of the temperature evolution with
 the analytic model, we estimated the progenitor size of SN 2024iss to be $R_{pro}=50-340$~$R_{\odot}$. 
 It is consistent with the range of Type IIb SNe.
 From these facts, we concluded that SN 2024iss 
 could be a core-collapse explosion from the 
 yellow (YSG) supergiant \citep{Morales-Garoffolo2014,Morales-Garoffolo2015,Tartaglia2017}. 

\begin{ack}
 We are grateful to graduate and undergraduate students for performing 
 the optical and near-infrared observations. This work was supported by Grant-in-Aid for Scientific Research (C) 22K03676. The Kagoshima University 1 m telescope is a member of the Optical and Infrared Synergetic Telescopes for Education and Research (OISTER) program funded by the MEXT of Japan.
\end{ack}

\bibliography{addsample}{}

\begin{thebibliography}{}
\expandafter\ifx\csname natexlab\endcsname\relax\def\natexlab#1{#1}\fi

\bibitem[{{Aldering} {et~al.}(1994){Aldering}, {Humphreys}, \&
  {Richmond}}]{Aldering1994}
{Aldering}, G., {Humphreys}, R.~M., \& {Richmond}, M. 1994, \aj, 107, 662

\bibitem[{{Arcavi} {et~al.}(2017){Arcavi}, {Hosseinzadeh}, {Brown}, {Smartt},
  {Valenti}, {Tartaglia}, {Piro}, {Sanchez}, {Nicholls}, {Monard}, {Howell},
  {McCully}, {Sand}, {Tonry}, {Denneau}, {Stalder}, {Heinze}, {Rest}, {Smith},
  \& {Bishop}}]{Arcavi2017}
{Arcavi}, I., {Hosseinzadeh}, G., {Brown}, P.~J., {et~al.} 2017, \apjl, 837, L2

\bibitem[{{Arnett}(1982)}]{Arnett1982}
{Arnett}, W.~D. 1982, \apj, 253, 785

\bibitem[{{Bellm} {et~al.}(2019){Bellm}, {Kulkarni}, {Graham}, {Dekany},
  {Smith}, {Riddle}, {Masci}, {Helou}, {Prince}, {Adams}, {Barbarino},
  {Barlow}, {Bauer}, {Beck}, {Belicki}, {Biswas}, {Blagorodnova}, {Bodewits},
  {Bolin}, {Brinnel}, {Brooke}, {Bue}, {Bulla}, {Burruss}, {Cenko}, {Chang},
  {Connolly}, {Coughlin}, {Cromer}, {Cunningham}, {De}, {Delacroix}, {Desai},
  {Duev}, {Eadie}, {Farnham}, {Feeney}, {Feindt}, {Flynn}, {Franckowiak},
  {Frederick}, {Fremling}, {Gal-Yam}, {Gezari}, {Giomi}, {Goldstein},
  {Golkhou}, {Goobar}, {Groom}, {Hacopians}, {Hale}, {Henning}, {Ho}, {Hover},
  {Howell}, {Hung}, {Huppenkothen}, {Imel}, {Ip}, {Ivezi{\'c}}, {Jackson},
  {Jones}, {Juric}, {Kasliwal}, {Kaspi}, {Kaye}, {Kelley}, {Kowalski},
  {Kramer}, {Kupfer}, {Landry}, {Laher}, {Lee}, {Lin}, {Lin}, {Lunnan},
  {Giomi}, {Mahabal}, {Mao}, {Miller}, {Monkewitz}, {Murphy}, {Ngeow},
  {Nordin}, {Nugent}, {Ofek}, {Patterson}, {Penprase}, {Porter}, {Rauch},
  {Rebbapragada}, {Reiley}, {Rigault}, {Rodriguez}, {van Roestel}, {Rusholme},
  {van Santen}, {Schulze}, {Shupe}, {Singer}, {Soumagnac}, {Stein}, {Surace},
  {Sollerman}, {Szkody}, {Taddia}, {Terek}, {Van Sistine}, {van Velzen},
  {Vestrand}, {Walters}, {Ward}, {Ye}, {Yu}, {Yan}, \& {Zolkower}}]{Bellm2019}
{Bellm}, E.~C., {Kulkarni}, S.~R., {Graham}, M.~J., {et~al.} 2019, \pasp, 131,
  018002

\bibitem[{{Bersten} {et~al.}(2012){Bersten}, {Benvenuto}, {Nomoto}, {Ergon},
  {Folatelli}, {Sollerman}, {Benetti}, {Botticella}, {Fraser}, {Kotak},
  {Maeda}, {Ochner}, \& {Tomasella}}]{Bersten2012}
{Bersten}, M.~C., {Benvenuto}, O.~G., {Nomoto}, K., {et~al.} 2012, \apj, 757,
  31

\bibitem[{{Bessell}(1990)}]{Bessell1990}
{Bessell}, M.~S. 1990, \pasp, 102, 1181

\bibitem[{{Bessell} {et~al.}(1998){Bessell}, {Castelli}, \&
  {Plez}}]{Bessell1998}
{Bessell}, M.~S., {Castelli}, F., \& {Plez}, B. 1998, \aap, 333, 231

\bibitem[{{Bianco} {et~al.}(2014){Bianco}, {Modjaz}, {Hicken}, {Friedman},
  {Kirshner}, {Bloom}, {Challis}, {Marion}, {Wood-Vasey}, \&
  {Rest}}]{Bianco2014}
{Bianco}, F.~B., {Modjaz}, M., {Hicken}, M., {et~al.} 2014, \apjs, 213, 19

\bibitem[{{Breeveld} {et~al.}(2011){Breeveld}, {Landsman}, {Holland}, {Roming},
  {Kuin}, \& {Page}}]{Breeveld2011}
{Breeveld}, A.~A., {Landsman}, W., {Holland}, S.~T., {et~al.} 2011, in American
  Institute of Physics Conference Series, Vol. 1358, Gamma Ray Bursts 2010, ed.
  J.~E. {McEnery}, J.~L. {Racusin}, \& N.~{Gehrels} (AIP), 373--376

\bibitem[{{Chambers} {et~al.}(2016){Chambers}, {Magnier}, {Metcalfe},
  {Flewelling}, {Huber}, {Waters}, {Denneau}, {Draper}, {Farrow}, {Finkbeiner},
  {Holmberg}, {Koppenhoefer}, {Price}, {Rest}, {Saglia}, {Schlafly}, {Smartt},
  {Sweeney}, {Wainscoat}, {Burgett}, {Chastel}, {Grav}, {Heasley}, {Hodapp},
  {Jedicke}, {Kaiser}, {Kudritzki}, {Luppino}, {Lupton}, {Monet}, {Morgan},
  {Onaka}, {Shiao}, {Stubbs}, {Tonry}, {White}, {Ba{\~n}ados}, {Bell},
  {Bender}, {Bernard}, {Boegner}, {Boffi}, {Botticella}, {Calamida},
  {Casertano}, {Chen}, {Chen}, {Cole}, {Deacon}, {Frenk}, {Fitzsimmons},
  {Gezari}, {Gibbs}, {Goessl}, {Goggia}, {Gourgue}, {Goldman}, {Grant},
  {Grebel}, {Hambly}, {Hasinger}, {Heavens}, {Heckman}, {Henderson}, {Henning},
  {Holman}, {Hopp}, {Ip}, {Isani}, {Jackson}, {Keyes}, {Koekemoer}, {Kotak},
  {Le}, {Liska}, {Long}, {Lucey}, {Liu}, {Martin}, {Masci}, {McLean}, {Mindel},
  {Misra}, {Morganson}, {Murphy}, {Obaika}, {Narayan}, {Nieto-Santisteban},
  {Norberg}, {Peacock}, {Pier}, {Postman}, {Primak}, {Rae}, {Rai}, {Riess},
  {Riffeser}, {Rix}, {R{\"o}ser}, {Russel}, {Rutz}, {Schilbach}, {Schultz},
  {Scolnic}, {Strolger}, {Szalay}, {Seitz}, {Small}, {Smith}, {Soderblom},
  {Taylor}, {Thomson}, {Taylor}, {Thakar}, {Thiel}, {Thilker}, {Unger},
  {Urata}, {Valenti}, {Wagner}, {Walder}, {Walter}, {Watters}, {Werner},
  {Wood-Vasey}, \& {Wyse}}]{Chambers2016}
{Chambers}, K.~C., {Magnier}, E.~A., {Metcalfe}, N., {et~al.} 2016, arXiv
  e-prints, arXiv:1612.05560

\bibitem[{{Crockett} {et~al.}(2008){Crockett}, {Eldridge}, {Smartt},
  {Pastorello}, {Gal-Yam}, {Fox}, {Leonard}, {Kasliwal}, {Mattila}, {Maund},
  {Stephens}, \& {Danziger}}]{Crockett2008}
{Crockett}, R.~M., {Eldridge}, J.~J., {Smartt}, S.~J., {et~al.} 2008, \mnras,
  391, L5

\bibitem[{{Cutri} {et~al.}(2003){Cutri}, {Skrutskie}, {van Dyk}, {Beichman},
  {Carpenter}, {Chester}, {Cambresy}, {Evans}, {Fowler}, {Gizis}, {Howard},
  {Huchra}, {Jarrett}, {Kopan}, {Kirkpatrick}, {Light}, {Marsh}, {McCallon},
  {Schneider}, {Stiening}, {Sykes}, {Weinberg}, {Wheaton}, {Wheelock}, \&
  {Zacarias}}]{Cutri2003}
{Cutri}, R.~M., {Skrutskie}, M.~F., {van Dyk}, S., {et~al.} 2003, VizieR Online
  Data Catalog, II/246

\bibitem[{{Ergon} {et~al.}(2014){Ergon}, {Sollerman}, {Fraser}, {Pastorello},
  {Taubenberger}, {Elias-Rosa}, {Bersten}, {Jerkstrand}, {Benetti},
  {Botticella}, {Fransson}, {Harutyunyan}, {Kotak}, {Smartt}, {Valenti},
  {Bufano}, {Cappellaro}, {Fiaschi}, {Howell}, {Kankare}, {Magill}, {Mattila},
  {Maund}, {Naves}, {Ochner}, {Ruiz}, {Smith}, {Tomasella}, \&
  {Turatto}}]{Ergon2014}
{Ergon}, M., {Sollerman}, J., {Fraser}, M., {et~al.} 2014, \aap, 562, A17

\bibitem[{{Filippenko} {et~al.}(1993){Filippenko}, {Matheson}, \&
  {Ho}}]{Filippenko1993}
{Filippenko}, A.~V., {Matheson}, T., \& {Ho}, L.~C. 1993, \apjl, 415, L103

\bibitem[{{Fukugita} {et~al.}(1996){Fukugita}, {Ichikawa}, {Gunn}, {Doi},
  {Shimasaku}, \& {Schneider}}]{Fukugita1996}
{Fukugita}, M., {Ichikawa}, T., {Gunn}, J.~E., {et~al.} 1996, \aj, 111, 1748

\bibitem[{{Godwin} {et~al.}(1983){Godwin}, {Metcalfe}, \& {Peach}}]{Godwin1983}
{Godwin}, J.~G., {Metcalfe}, N., \& {Peach}, J.~V. 1983, \mnras, 202, 113

\bibitem[{{Kumar} {et~al.}(2013){Kumar}, {Pandey}, {Sahu}, {Vinko},
  {Moskvitin}, {Anupama}, {Bhatt}, {Ordasi}, {Nagy}, {Sokolov}, {Sokolova},
  {Komarova}, {Kumar}, {Bose}, {Roy}, \& {Sagar}}]{Kumar2013}
{Kumar}, B., {Pandey}, S.~B., {Sahu}, D.~K., {et~al.} 2013, \mnras, 431, 308

\bibitem[{{Lyman} {et~al.}(2016){Lyman}, {Bersier}, {James}, {Mazzali},
  {Eldridge}, {Fraser}, \& {Pian}}]{Lyman2016}
{Lyman}, J.~D., {Bersier}, D., {James}, P.~A., {et~al.} 2016, \mnras, 457, 328

\bibitem[{{Marion} {et~al.}(2014){Marion}, {Vinko}, {Kirshner}, {Foley},
  {Berlind}, {Bieryla}, {Bloom}, {Calkins}, {Challis}, {Chevalier}, {Chornock},
  {Culliton}, {Curtis}, {Esquerdo}, {Everett}, {Falco}, {France}, {Fransson},
  {Friedman}, {Garnavich}, {Leibundgut}, {Meyer}, {Smith}, {Soderberg},
  {Sollerman}, {Starr}, {Szklenar}, {Takats}, \& {Wheeler}}]{Marion2014}
{Marion}, G.~H., {Vinko}, J., {Kirshner}, R.~P., {et~al.} 2014, \apj, 781, 69

\bibitem[{{Medler} {et~al.}(2022){Medler}, {Mazzali}, {Teffs}, {Ashall},
  {Anderson}, {Arcavi}, {Benetti}, {Bostroem}, {Burke}, {Cai},
  {Charalampopoulos}, {Elias-Rosa}, {Ergon}, {Galbany}, {Gromadzki},
  {Hiramatsu}, {Howell}, {Inserra}, {Lundqvist}, {McCully}, {M{\"u}ller-Bravo},
  {Newsome}, {Nicholl}, {Padilla Gonzalez}, {Paraskeva}, {Pastorello},
  {Pellegrino}, {Pessi}, {Reguitti}, {Reynolds}, {Roy}, {Terreran},
  {Tomasella}, \& {Young}}]{Medler2022}
{Medler}, K., {Mazzali}, P.~A., {Teffs}, J., {et~al.} 2022, \mnras, 513, 5540

\bibitem[{{Milisavljevic} {et~al.}(2013){Milisavljevic}, {Margutti},
  {Soderberg}, {Pignata}, {Chomiuk}, {Fesen}, {Bufano}, {Sanders}, {Parrent},
  {Parker}, {Mazzali}, {Pian}, {Pickering}, {Buckley}, {Crawford}, {Gulbis},
  {Hettlage}, {Hooper}, {Nordsieck}, {O'Donoghue}, {Husser}, {Potter},
  {Kniazev}, {Kotze}, {Romero-Colmenero}, {Vaisanen}, {Wolf}, {Bietenholz},
  {Bartel}, {Fransson}, {Walker}, {Brunthaler}, {Chakraborti}, {Levesque},
  {MacFadyen}, {Drescher}, {Bock}, {Marples}, {Anderson}, {Benetti},
  {Reichart}, \& {Ivarsen}}]{Milisavljevic2013}
{Milisavljevic}, D., {Margutti}, R., {Soderberg}, A.~M., {et~al.} 2013, \apj,
  767, 71

\bibitem[{{Morales-Garoffolo} {et~al.}(2014){Morales-Garoffolo}, {Elias-Rosa},
  {Benetti}, {Taubenberger}, {Cappellaro}, {Pastorello}, {Klauser}, {Valenti},
  {Howerton}, {Ochner}, {Schramm}, {Siviero}, {Tartaglia}, \&
  {Tomasella}}]{Morales-Garoffolo2014}
{Morales-Garoffolo}, A., {Elias-Rosa}, N., {Benetti}, S., {et~al.} 2014,
  \mnras, 445, 1647

\bibitem[{{Morales-Garoffolo} {et~al.}(2015){Morales-Garoffolo}, {Elias-Rosa},
  {Bersten}, {Jerkstrand}, {Taubenberger}, {Benetti}, {Cappellaro}, {Kotak},
  {Pastorello}, {Bufano}, {Dom{\'\i}nguez}, {Ergon}, {Fraser}, {Gao},
  {Garc{\'\i}a}, {Howell}, {Isern}, {Smartt}, {Tomasella}, \&
  {Valenti}}]{Morales-Garoffolo2015}
{Morales-Garoffolo}, A., {Elias-Rosa}, N., {Bersten}, M., {et~al.} 2015,
  \mnras, 454, 95

\bibitem[{{Nagayama} \& {Nakaya}(2024)}]{Nagayama2024}
{Nagayama}, T., \& {Nakaya}, H. 2024, in Society of Photo-Optical
  Instrumentation Engineers (SPIE) Conference Series, Vol. 13096, Ground-based
  and Airborne Instrumentation for Astronomy X, ed. J.~J. {Bryant},
  K.~{Motohara}, \& J.~R.~D. {Vernet}, 130963I

\bibitem[{{Nakar} \& {Sari}(2010)}]{NS2010}
{Nakar}, E., \& {Sari}, R. 2010, \apj, 725, 904

\bibitem[{{Nomoto} {et~al.}(1993){Nomoto}, {Suzuki}, {Shigeyama}, {Kumagai},
  {Yamaoka}, \& {Saio}}]{Nomoto1993}
{Nomoto}, K., {Suzuki}, T., {Shigeyama}, T., {et~al.} 1993, \nat, 364, 507

\bibitem[{{O'Neill} {et~al.}(2024){O'Neill}, {Godson}, {Killestein}, {Kotak},
  {Kuncarayakti}, {Pursiainen}, {Ackley}, {Dyer}, {Jim{\'e}nez-Ibarra},
  {Lyman}, {Ulaczyk}, {Steeghs}, {Galloway}, {Dhillon}, {O'Brien}, {Ramsay},
  {Noysena}, {Breton}, {Nuttall}, {Pall{\'e}}, {Pollacco}, \&
  {Kumar}}]{ONeill2024}
{O'Neill}, D., {Godson}, B., {Killestein}, T., {et~al.} 2024, Transient Name
  Server AstroNote, 128, 1

\bibitem[{{Pastorello} {et~al.}(2008){Pastorello}, {Kasliwal}, {Crockett},
  {Valenti}, {Arbour}, {Itagaki}, {Kaspi}, {Gal-Yam}, {Smartt}, {Griffith},
  {Maguire}, {Ofek}, {Seymour}, {Stern}, \& {Wiethoff}}]{Pastorello2008}
{Pastorello}, A., {Kasliwal}, M.~M., {Crockett}, R.~M., {et~al.} 2008, \mnras,
  389, 955

\bibitem[{{Pellegrino} {et~al.}(2023){Pellegrino}, {Hiramatsu}, {Arcavi},
  {Howell}, {Bostroem}, {Brown}, {Burke}, {Elias-Rosa}, {Itagaki}, {Kaneda},
  {McCully}, {Modjaz}, {Padilla Gonzalez}, {Pritchard}, \&
  {Yesmin}}]{Pellegrino2023}
{Pellegrino}, C., {Hiramatsu}, D., {Arcavi}, I., {et~al.} 2023, \apj, 954, 35

\bibitem[{{P{\'e}rez-Fournon} {et~al.}(2024){P{\'e}rez-Fournon}, {Poidevin},
  {Aguado}, {Acosta-Pulido}, {L{\'o}pez-Oramas}, {Nespral}, \&
  {Acero}}]{PFournon2024}
{P{\'e}rez-Fournon}, I., {Poidevin}, F., {Aguado}, D.~S., {et~al.} 2024,
  Transient Name Server AstroNote, 130, 1

\bibitem[{{Richmond} {et~al.}(1994){Richmond}, {Treffers}, {Filippenko},
  {Paik}, {Leibundgut}, {Schulman}, \& {Cox}}]{Richmond1994}
{Richmond}, M.~W., {Treffers}, R.~R., {Filippenko}, A.~V., {et~al.} 1994, \aj,
  107, 1022

\bibitem[{{Roming} {et~al.}(2005){Roming}, {Kennedy}, {Mason}, {Nousek}, {Ahr},
  {Bingham}, {Broos}, {Carter}, {Hancock}, {Huckle}, {Hunsberger}, {Kawakami},
  {Killough}, {Koch}, {McLelland}, {Smith}, {Smith}, {Soto}, {Boyd},
  {Breeveld}, {Holland}, {Ivanushkina}, {Pryzby}, {Still}, \&
  {Stock}}]{Roming2005}
{Roming}, P. W.~A., {Kennedy}, T.~E., {Mason}, K.~O., {et~al.} 2005, \ssr, 120,
  95

\bibitem[{{Sana} {et~al.}(2012){Sana}, {de Mink}, {de Koter}, {Langer},
  {Evans}, {Gieles}, {Gosset}, {Izzard}, {Le Bouquin}, \&
  {Schneider}}]{Sana2012}
{Sana}, H., {de Mink}, S.~E., {de Koter}, A., {et~al.} 2012, Science, 337, 444

\bibitem[{{S{\'a}nchez-S{\'a}ez} {et~al.}(2021){S{\'a}nchez-S{\'a}ez}, {Reyes},
  {Valenzuela}, {F{\"o}rster}, {Eyheramendy}, {Elorrieta}, {Bauer},
  {Cabrera-Vives}, {Est{\'e}vez}, {Catelan}, {Pignata}, {Huijse}, {De Cicco},
  {Ar{\'e}valo}, {Carrasco-Davis}, {Abril}, {Kurtev}, {Borissova}, {Arredondo},
  {Castillo-Navarrete}, {Rodriguez}, {Ruz-Mieres}, {Moya},
  {Sabatini-Gacit{\'u}a}, {Sep{\'u}lveda-Cobo}, \&
  {Camacho-I{\~n}iguez}}]{ZTF2021}
{S{\'a}nchez-S{\'a}ez}, P., {Reyes}, I., {Valenzuela}, C., {et~al.} 2021, \aj,
  161, 141

\bibitem[{{Schlafly} \& {Finkbeiner}(2011)}]{Schlafly2011}
{Schlafly}, E.~F., \& {Finkbeiner}, D.~P. 2011, \apj, 737, 103

\bibitem[{{Smith} {et~al.}(2011){Smith}, {Li}, {Filippenko}, \&
  {Chornock}}]{Smith2011}
{Smith}, N., {Li}, W., {Filippenko}, A.~V., \& {Chornock}, R. 2011, \mnras,
  412, 1522

\bibitem[{{Srivastav} {et~al.}(2024{\natexlab{a}}){Srivastav}, {Fulton},
  {Nicholl}, {Angus}, {Smith}, {Young}, {Moore}, {Sim}, {Smartt}, \&
  {Chen}}]{Srivastav2024a}
{Srivastav}, S., {Fulton}, M., {Nicholl}, M., {et~al.} 2024{\natexlab{a}},
  Transient Name Server Classification Report, 2024-1517, 1

\bibitem[{{Srivastav} {et~al.}(2024{\natexlab{b}}){Srivastav}, {Fulton},
  {Nicholl}, {Angus}, {Smith}, {Young}, {Moore}, {Sim}, {Smartt}, \&
  {Chen}}]{Srivastav2024b}
---. 2024{\natexlab{b}}, Transient Name Server Classification Report,
  2024-2046, 1

\bibitem[{{Steeghs} {et~al.}(2022){Steeghs}, {Galloway}, {Ackley}, {Dyer},
  {Lyman}, {Ulaczyk}, {Cutter}, {Mong}, {Dhillon}, {O'Brien}, {Ramsay},
  {Poshyachinda}, {Kotak}, {Nuttall}, {Pall{\'e}}, {Breton}, {Pollacco},
  {Thrane}, {Aukkaravittayapun}, {Awiphan}, {Burhanudin}, {Chote}, {Chrimes},
  {Daw}, {Duffy}, {Eyles-Ferris}, {Gompertz}, {Heikkil{\"a}}, {Irawati},
  {Kennedy}, {Killestein}, {Kuncarayakti}, {Levan}, {Littlefair},
  {Makrygianni}, {Marsh}, {Mata-Sanchez}, {Mattila}, {Maund}, {McCormac},
  {Mkrtichian}, {Mullaney}, {Noysena}, {Patel}, {Rol}, {Sawangwit}, {Stanway},
  {Starling}, {Str{\o}m}, {Tooke}, {West}, {White}, \&
  {Wiersema}}]{Steeghs2022}
{Steeghs}, D., {Galloway}, D.~K., {Ackley}, K., {et~al.} 2022, \mnras, 511,
  2405

\bibitem[{{Stetson}(1987)}]{Stetson1987}
{Stetson}, P.~B. 1987, \pasp, 99, 191

\bibitem[{{Szalai} {et~al.}(2016){Szalai}, {Vink{\'o}}, {Nagy}, {Silverman},
  {Wheeler}, {Dhungana}, {Marion}, {Kehoe}, {Fox}, {S{\'a}rneczky},
  {Marschalk{\'o}}, {B{\'\i}r{\'o}}, {Borkovits}, {Heged{\"u}s}, {Szak{\'a}ts},
  {Ferrante}, {B{\'a}nyai}, {Hodos{\'a}n}, {Kelemen}, \&
  {P{\'a}l}}]{Szalai2016}
{Szalai}, T., {Vink{\'o}}, J., {Nagy}, A.~P., {et~al.} 2016, \mnras, 460, 1500

\bibitem[{{Tartaglia} {et~al.}(2017){Tartaglia}, {Fraser}, {Sand}, {Valenti},
  {Smartt}, {McCully}, {Anderson}, {Arcavi}, {Elias-Rosa}, {Galbany},
  {Gal-Yam}, {Haislip}, {Hosseinzadeh}, {Howell}, {Inserra}, {Jha}, {Kankare},
  {Lundqvist}, {Maguire}, {Mattila}, {Reichart}, {Smith}, {Smith},
  {Stritzinger}, {Sullivan}, {Taddia}, \& {Tomasella}}]{Tartaglia2017}
{Tartaglia}, L., {Fraser}, M., {Sand}, D.~J., {et~al.} 2017, \apjl, 836, L12

\bibitem[{{Tody}(1986)}]{Tody1986}
{Tody}, D. 1986, in Society of Photo-Optical Instrumentation Engineers (SPIE)
  Conference Series, Vol. 627, Instrumentation in astronomy VI, ed. D.~L.
  {Crawford}, 733

\bibitem[{{Tody}(1993)}]{Tody1993}
{Tody}, D. 1993, in Astronomical Society of the Pacific Conference Series,
  Vol.~52, Astronomical Data Analysis Software and Systems II, ed. R.~J.
  {Hanisch}, R.~J.~V. {Brissenden}, \& J.~{Barnes}, 173

\bibitem[{{Tokunaga} \& {Vacca}(2005)}]{Tokunaga2005}
{Tokunaga}, A.~T., \& {Vacca}, W.~D. 2005, \pasp, 117, 421

\bibitem[{{Van Dyk} {et~al.}(2014){Van Dyk}, {Zheng}, {Fox}, {Cenko}, {Clubb},
  {Filippenko}, {Foley}, {Miller}, {Smith}, {Kelly}, {Lee}, {Ben-Ami}, \&
  {Gal-Yam}}]{VanDyk2014}
{Van Dyk}, S.~D., {Zheng}, W., {Fox}, O.~D., {et~al.} 2014, \aj, 147, 37

\bibitem[{{Woosley} {et~al.}(1994){Woosley}, {Eastman}, {Weaver}, \&
  {Pinto}}]{Woosley1994}
{Woosley}, S.~E., {Eastman}, R.~G., {Weaver}, T.~A., \& {Pinto}, P.~A. 1994,
  \apj, 429, 300

\end{thebibliography}
\bibliographystyle{apj}

\end{document}